# NOISE MITIGATION METHODS FOR DIGITAL VISIBLE LIGHT COMMUNICATION

Wataru Uemura [1] and Takumi Hamano [2]

[1] Faculty of Advanced Science and Technology, Ryukoku University, Shiga, Japan
[2] Faculty of Science and Technology, Ryukoku University, Shiga, Japan

## ABSTRACT

*Visible Light Communication (VLC) using Light Emitting Diodes (LEDs) has gained attention due to its low power consumption, long lifetime, and fast response. However, VLC suffers from optical noise generated by ambient light sources such as fluorescent lamps, which leads to waveform distortion and increased bit error rates (BER). In this paper, we propose two noise reduction methods for Digital Visible Light Communication (DVLC) systems. The first method exploits the periodic nature of interference caused by AC-powered-line illumination and reduces interference by subtracting sampled noise waveforms from the received signal. Second, inspired by Active Noise Control (ANC) techniques, an additional photodiode is introduced for noise reception, and subtraction circuits are employed to attenuate noise in real time. Experimental results show that both methods improve BER performance compared with conventional receivers, with the ANC-inspired approach achieving superior performance under all tested conditions.*

## 1. INTRODUCTION

Visible Light Communication (VLC) has emerged as promising wireless communication technology that exploits the visible spectrum between 380 nm and 780 nm. Unlike radio frequency (RF) systems, VLC provides clearly defined communication areas determined by illumination and can be seamlessly integrated into existing lighting infrastructure. However, the performance of VLC can be significantly degraded by optical noise originating from ambient light sources such as fluorescent lamps [1 - 5]. Digital Visible Light Communication (DVLC) refers to VLC systems that employ digital modulation schemes such as OOK, PPM, or I-PPM to transmit binary data through visible light.

In DVLC systems, the receiver typically distinguishes HIGH and LOW states of the signal by applying a threshold to the received voltage. When optical noise is superimposed on the signal, this thresholding process can lead to incorrect symbol detection, thereby degrading communication reliability.

In this paper, we address the problem of optical noise in VLC and propose two methods that enable robust demodulation in noisy environments. Section 2 provides an overview of the fundamentals of VLC, while Section 3 discusses the problems caused by noise in VLC. Two general categories of noise are considered: periodic and non-periodic. In Section 4, we focus on periodic noise caused by power-line interference, which depends on the frequency of the AC power supply (typically 50 Hz or 60 Hz worldwide). For example, in western Japan the frequency is 60 Hz, and all experiments in this paper are conducted under such conditions. To mitigate this interference, we propose a subtraction method in which one cycle of the noise waveform is sampled and removed from the received signal, thereby reducing the noise amplitude.

In Section 5, we focus on non-periodic noise. An auxiliary photodetector is introduced to capture only the noise component, excluding the VLC signal. This reference noise signal is then subtracted from the received waveform in real time using a subtraction circuit.

The effectiveness of both proposed methods is evaluated by comparing the relationship between the energy-per-bit-to-noise ratio ($E_b/N_0$) and the bit error rate (BER) against a conventional method without noise mitigation. Finally, Section 6 presents the conclusions of the paper.

## 2. VISIBLE LIGHT COMMUNICATION (VLC)

Visible Light Communication (VLC) is a wireless communication technology that uses electromagnetic waves in the visible spectrum, with wavelengths ranging from 380 nm to 780 nm [6 - 10]. In practical systems, white light-emitting diodes (LEDs) are commonly used as transmitters, either by combining multiple wavelengths across the visible band or by mixing blue and yellow light. In digital VLC using pulse modulation, the ON state of the light source is assigned to HIGH, while the OFF state corresponds to LOW. Information is conveyed through a sequence of time slots, with each slot represented as either HIGH or LOW.

LEDs have been extensively investigated as transmitters in VLC. Compared with conventional lighting devices such as incandescent bulbs, LEDs provide distinct advantages: lower power consumption, longer operational lifetime, and much faster response speed. This rapid response enables high data rates in VLC. Furthermore, their ability to switch ON and OFF at high frequencies effectively suppresses perceptible flicker, allowing LEDs to function simultaneously as illumination and communication devices.

On the receiver side, photodiodes are widely used as detectors because their response time is faster than that of phototransistors. Since VLC employs light as the transmission carrier, ordinary illumination devices can also act as transmitters. Moreover, because communication is confined strictly to the illuminated area, the coverage of VLC is well-defined, in contrast to radio frequency (RF) communication, where the boundaries of the communication range are less distinct. This spatial confinement also provides security advantages, as eavesdropping from outside the illuminated area is inherently difficult [11, 12]. Table 1 summarizes the features of VLC.

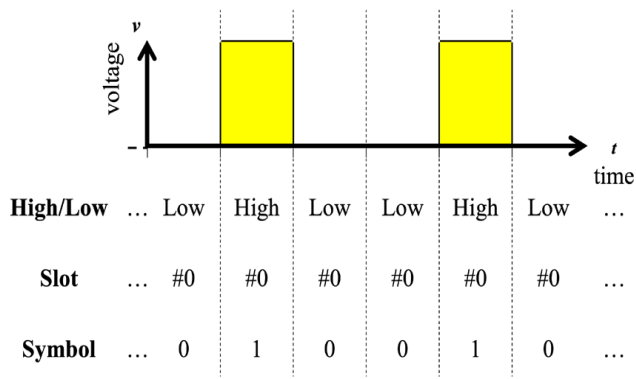

Figure 1: On-Off keying which assigns the HIGH state to "1" and the LOW state to "0"

International Journal of Computer Networks & Communications (IJCNC) Vol.18, No.1, January 2026

Table 1  Features of Visible Light Communication (VLC)

| Advantages | Disadvantages |
| --- | --- |
| High data rate enabled by fast LED response | Strongly affected by ambient light noise |
| Energy efficient and long lifetime of LEDs | Requires line-of-sight or sufficient illumination |
| No radio frequency interference | Limited coverage compared to RF communication |
| Clearly defined communication area | Coverage depends on lighting conditions |
| Security benefits due to confined communication space [11, 12] | Uplink communication is difficult |
| Integration with illumination (dual-use for lighting and communication) | Flicker and dimming must be carefully controlled |

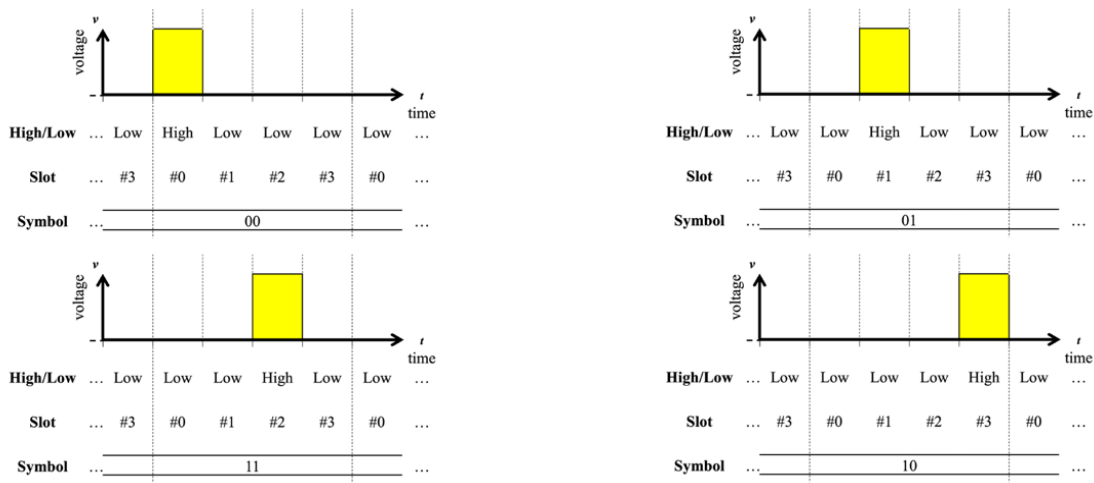

Figure 2: An example where four slots are used for 4PPM.

## 2.1. Flicker in VLC

Flicker occurs when the duty ratio of HIGH and LOW states within one cycle of a transmission slot is not constant. In such cases, the human eye perceives rapid and irregular fluctuations in brightness, rather than simple high-speed ON-OFF switching. Continuous exposure to flickering light can have adverse effects on human health, such as dizziness and nausea. Therefore, in VLC systems where lighting devices are also used as transmitters, modulation methods must be carefully designed to avoid flicker.

## 2.2. Modulation Methods for VLC

This subsection introduces several modulation schemes for VLC: On-Off Keying (OOK), Pulse Width Modulation (PWM), Pulse Position Modulation (PPM), and Inversed Pulse Position Modulation (I-PPM). Among these, OOK and PWM are prone to flicker and are therefore not suitable when VLC is used simultaneously for illumination. In contrast, PPM and I-PPM maintain a constant duty ratio within each cycle, thereby preventing flicker. For this reason, these two modulations have been the focus of much recent VLC research.

### 2.2.1. On-Off Keying (OOK)

As shown in Figure 1, OOK transmits binary data by assigning the HIGH state to "1" and the

53

LOW state to "0" (yellow indicates HIGH and white indicates LOW in the figure). OOK is the simplest digital modulation scheme to implement. However, since it does not maintain a constant duty ratio, flicker can occur. Consequently, OOK is generally unsuitable for VLC systems that rely on lighting devices as transmitters.

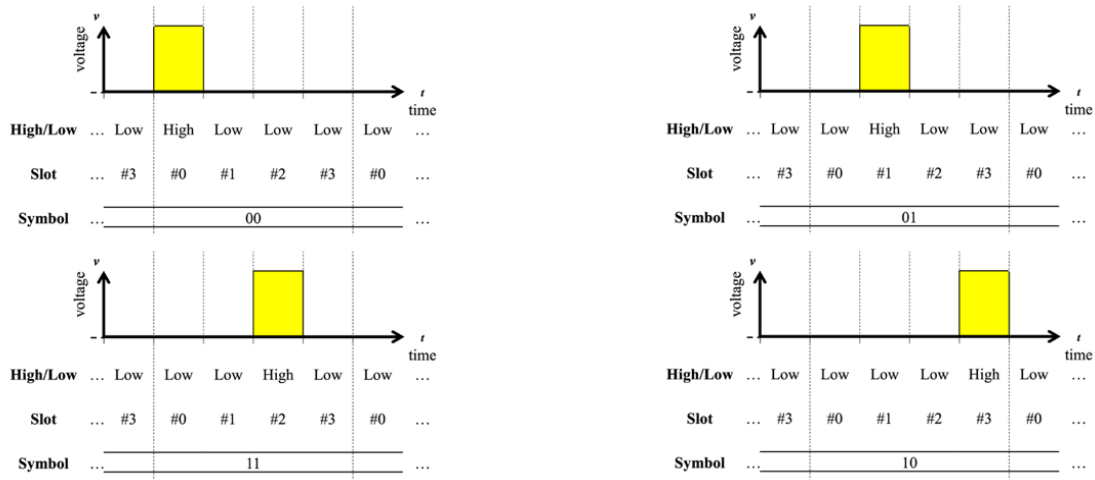

Figure 3: An example where four slots are used for I-4PPM.

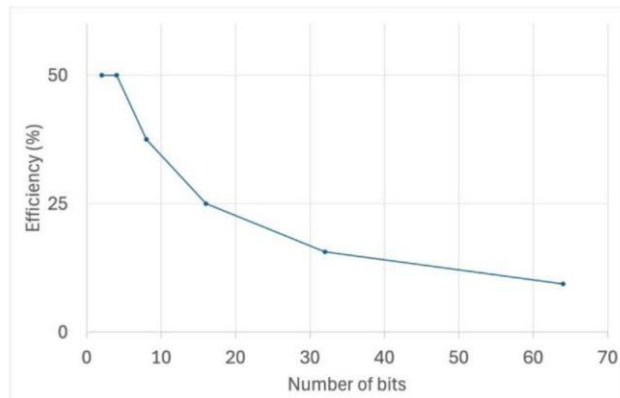

Figure 4: The relationship between the brightness (number of bits) and the transmission efficiency.

### 2.2.2. Pulse Position Modulation (PPM)

As shown in Figure 2, PPM encodes information by changing the position of a pulse within a predefined number of slots [13]. The example in the figure uses four slots. In PPM, the duty ratio is constant across all symbol assignments, so flicker is inherently avoided. For this reason, PPM is suitable for VLC systems.

However, the ratio of HIGH-to-LOW is fixed at 1:$N$ (where $N$ is the number of slots minus one). As the number of slots increases, the proportion of LOW states becomes larger, making the illumination appear dimmer to the human eye. Furthermore, increasing the number of slots reduces transmission efficiency, since the maximum duty ratio cannot exceed 50%. Therefore, when VLC transmitters are also used for illumination, PPM may not be the most practical choice.

### 2.2.3. Inverse Pulse Position Modulation (I-PPM)

As depicted in Figure 3, I-PPM encodes binary information by shifting the position of the LOW state within a fixed number of slots. While conventional PPM which uses the HIGH state as the information carrier, I-PPM instead assigns information based on the LOW state. Figure 3 shows an example with four slots. Like PPM, I-PPM maintains a constant duty ratio and thus voids flicker, making it well-suited for VLC.

In contrast to PPM, the HIGH-to-LOW ratio in I-PPM is $N$:1 (where $N$ is the number of slots minus one). As the slot count increases, the position of HIGH states becomes larger, producing brighter illumination. For this reason, I-PPM is often preferred in VLC systems where both communication and lighting requirements must be satisfied.

The relationship between the number of bits and the number of slots in I-PPM is expressed in Eq. (1), and the corresponding transmission efficiency is given in Eq. (2). The calculated values are summarized in and illustrated in Figure 4.

Table 2: Information transmission efficiency of I-PPM

| **Number of slots** | 1 | 2 | 3 | 4 | 5 | 6 |
|---|---|---|---|---|---|---|
| Number of bits | 2 | 4 | 8 | 16 | 32 | 64 |
| Duty ratio (%) | 50.0 | 75.0 | 87.5 | 93.8 | 96.9 | 98.4 |
| Efficiency (%) | 50.0 | 50.0 | 37.5 | 25.0 | 15.6 | 9.4 |

$$\text{Number of bits} = 2^S \quad (S : \text{number of slots}) \quad (1)$$
$$\text{Transmission efficiency} = 2^S / S \quad (2)$$

As shown in Figure 4, increasing the number of slots —thereby producing brighter illumination—inevitably reduces transmission efficiency. A common countermeasure is to shorten the symbol duration, which makes it possible to increase the slot count without sacrificing efficiency. However, as the modulation order increases, although the communication rate improves, the time allocated per bit becomes shorter, making the system increasingly susceptible to noise. Therefore, the slot number must be carefully managed with respect to both transmission efficiency and robustness against noise.

### 2.3. Digital Visible Light Communication

#### 2.3.1. Transmitter

Transmission process begins with data provided from a terminal device, such as a computer, to a microcontroller that performs modulation. Based on the input data, the microcontroller devices the LED to emit modulated optical signals. The specific modulation scheme is predetermined by the program implemented on the microcontroller.

#### 2.3.2. Receiver

On the receiver side, a photodiode converts incident light into corresponding voltage variations. These signals are classified into HIGH and LOW levels using a comparator, which functions as a specialized operational amplifier. The comparator determines the logic level by comparing the

International Journal of Computer Networks & Communications (IJCNC) Vol.18, No.1, January 2026

voltage at its non-inverting input with a reference threshold applied to its inverting input. If non-inverting voltage exceeds the threshold, the comparator outputs the positive supply voltage, interpreted by the microcontroller as HIGH; otherwise, it outputs the negative supply voltage, interpreted as LOW. The resulting sequence of HIGH and LOW signals is then demodulated by the microcontroller, according to its programmed algorithm, and the recovered information is transferred back to a terminal device such as a computer.

## 3. NOISE IN DIGITAL VISIBLE LIGHT COMMUNICATION

This section discusses the challenges posed by noise in digital VLC systems and explains its sources and effects.

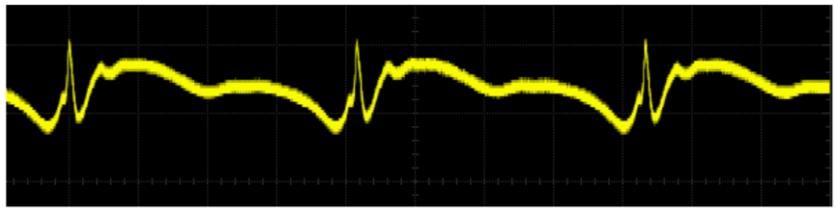

Figure 5: Fluorescent lamps generate noise waveforms with distortions.

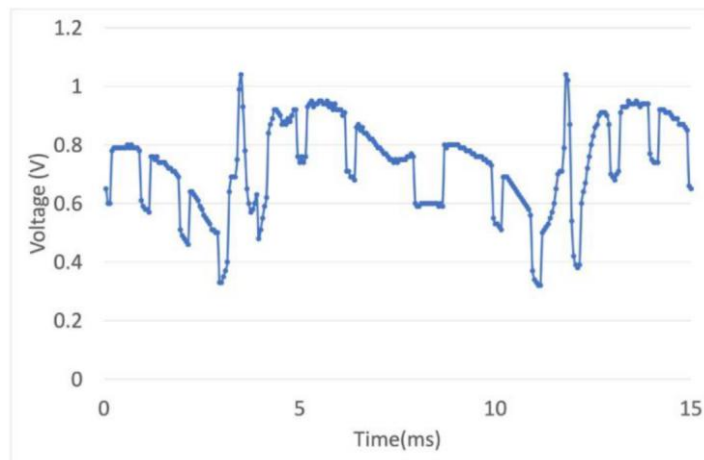

Figure 6: When the noise is superimposed on a digital

### 3.1. Overview of Noise

In digital VLC, light sources other than the intended transmitter act as interference. A typical example is fluorescent lighting, which produces distorted noise waveforms, as illustrated in Figure 5. When such noise is superimposed on a digital VLC signal, the received waveform becomes distorted, as illustrated in Figure 6. The origin of this distortion lies in the rectification process of alternating current (AC) power, which is examined in the next subsection.

### 3.2. Rectification of AC Voltage

Rectification is the process of converting alternating current (AC) into direct current (DC), performed by a rectifier circuit. A common implementation is the diode-based full-wave or bridge rectifier, which makes use of both the positive and negative halves of the AC cycle. In such circuit

56

to produce a quasi-DC waveform. However, the resulting signal is not perfectly constant due to transient phenomena. In addition, the frequency of the rectified output becomes twice that of the original AC input.

### 3.3. Transient Phenomena and Signal Distortion

In VLC receivers, HIGH and LOW states are determined relative to a threshold voltage. In Figure 7 illustrates an example in which the threshold (red line) is set near the center of the signal shown in Figure 6. As indicated, the segment marked with a red circle should represent a LOW signal, but noise raises it above the threshold, leading to a false HIGH detection. Conversely, the segment marked with a green circle should represent a HIGH signal, but noise effectively elevates the threshold, causing it to be misinterpreted as LOW.

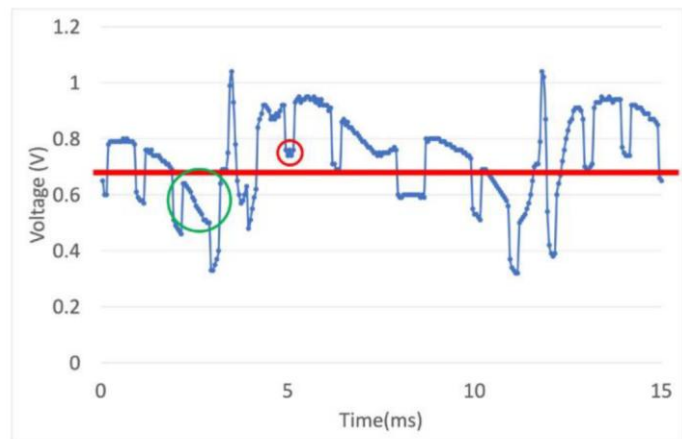

Figure 7: A threshold (red line) is drawn approximately at the centre of the signal level.

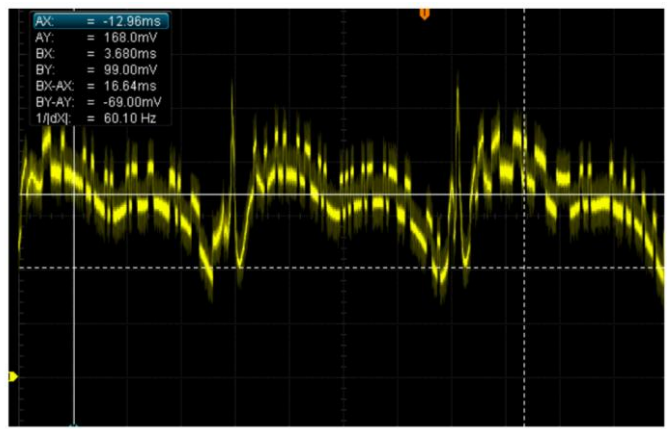

Figure 8: A received waveform when the communication speed is 10 kbps.

This misclassification becomes increasingly severe as the communication speed rises. For example, Figure 8 shows a 10 kbps waveform, while Figure 9 shows one at 2.5 kbps. In the presence of fluorescent lamp noise, more symbols fall within a single noise cycle at higher transmission rates, increasing the probability of detection errors. These observations indicate that the number of misclassifications grows proportionally with communication speed.

### 3.4. Impact of Noise on VLC

Transient phenomena generally occur in circuits containing inductors or capacitors during transitions between steady states. A steady state is defined as a condition in which circuit variables such as voltage remain constant over time or exhibit periodic behavior synchronized with the input frequency. Furthermore, the distortion observed in fluorescent lamp waveforms, as shown in Figure 5, is caused by the fact that the rectified AC voltage is derived from commercial power supplies operating at very high voltages.

### 3.5. Filter Circuits

One conventional approach for noise reduction in VLC is the application of filter circuits. A filter circuit selectively extracts the desired frequency components from an input signal that contains multiple frequency components. Representative examples are described below.

#### 3.5.1. Low-Pass Filter

A low-pass filter attenuates signal components above a certain cutoff frequency while allowing lower-frequency components to pass. It is also referred to as a high-cut filter or low-frequency

$$|G(j\omega)| = \frac{1}{\sqrt{1+(\omega CR)^2}} \quad (3)$$

$$f_c = \frac{1}{2\pi CR} \quad (4)$$

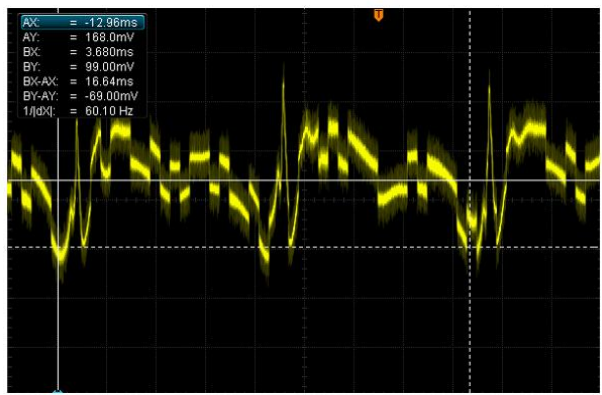

Figure 9: A received waveform when the communication speed is 2.5 kbps.

pass filter and is commonly used in devices such as digital cameras. A first-order RC low-pass filter, consisting of a resistor and a capacitor, exhibits the gain shown in Eq. (3), where $\omega$ denotes angular frequency. The cutoff frequency $fc$ is given by Eq. (4).

### 3.5.2. High-Pass Filter

A high-pass filter suppresses low-frequency signals components while allowing high-frequency components to pass. It is also called a low-cut filter or high-frequency pass filter and is widely used in audio signal processing. The gain of a first-order RC high-pass filter is expressed in Eq. (5), while the cutoff frequency $fc$ is determined by the same expression as in Eq. (4).

$$|G(j\omega)| = \frac{\omega CR}{\sqrt{1 + (\omega CR)^2}} \quad (5)$$

### 3.5.3. Band-Pass Filter

A band-pass filter allows only a specified frequency band to pass while attenuating frequencies outside that range. Also referred to as a band-pass network, it is commonly used in communication devices such as smartphones and wireless transceivers. The gain ($\omega$) of a parallel LC resonance circuit combined with a resistor is shown in Eq. (6), and the resonance frequency $f_0$ is in Eq. (7).

$$|G(\omega)| = \frac{\omega CR}{1 + (\omega CR)^2} \quad (6)$$

$$f_0 = \frac{1}{2\pi\sqrt{LC}} \quad (7)$$

### 3.5.4. Discussion

In digital VLC systems, filter circuits are an effective means of extracting only the frequency components relevant to signal reception. However, their effectiveness is fundamentally limited: if the noise frequency is close to or overlaps with the signal frequency, filtering alone cannot sufficiently suppress the interference. The limitation motivates the development of alternative noise mitigation methods, which are discussed in the following sections.

## 4. NOISE MITIGATION METHODS IN VISIBLE LIGHT COMMUNICATION

In this section, we propose solutions to address the problems that arise in indoor digital VLC systems, particularly the effect of interference caused by ambient light sources such as fluorescent lamps.

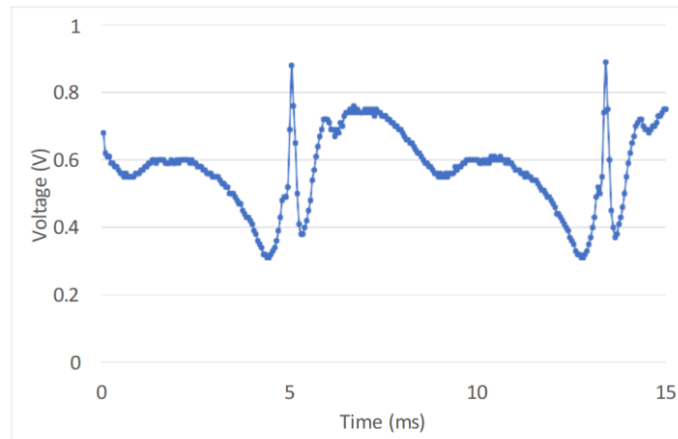

Figure 10: An example of the noise waveform observed in our experiments.

Table 3: Procedure of the proposed method

| step | Description |
| --- | --- |
| 1 | Sample one full cycle of the interference waveform during idle periods using an AD converter and store it as the reference subtraction value. |
| 2 | For subsequent reception, subtract the stored interference value from the incoming noisy signal. |
| 3 | Output the noise-reduced signal to the demodulation microcontroller through a DA converter. |
| 4 | Demodulate the signal using the microcontroller and transfer the recovered information to the PC. |

## 4.1. Periodic Interference Subtraction Method

In indoor VLC environments, the received signal can be significantly distorted by optical noise from other light sources, especially fluorescent lamps. Figure 10 shows an example of the noise waveform observed in our experiments. As explained in Section 3, this distortion originates from the transient phenomenon caused by rectifying the commercial AC power supply into DC. Since the source is AC, the resulting interference exhibits strong periodicity, as illustrated in Figure 10. Based on this observation, we propose a method that mitigates the effect of such interference by using the periodic characteristics of the waveform.

At the receiver, both the interference signal and the received VLC signal are sampled as analog values through an analog-to-digital (AD) converter. One cycle of the periodic interference waveform is then subtracted from the received signal waveform. The resulting noise-reduced signal is converted back into analog form using a digital-to-analog (DA) converter and passed to the demodulation microcontroller, which recovers the transmitted information. During the sampling process, the interval is set to capture five points per slot of the received VLC signal. This design prevents the distortion that would occur if the sharp edges of rectangular waveforms are sampled directly. The procedure of the proposed method is summarized in Table 3.

## 4.2. Implementation

We implement the proposed system and validate its functionality. In this experiment, the objective is to verify the behavior of the signal processing. Therefore, transmission data are preloaded into the modulation microcontroller, while the demodulation microcontroller is used for both demodulation and data reception. The main components of the system are summarized below.

**Transmitter:**

- Microcontroller: Arduino Nano (ATmega328P)

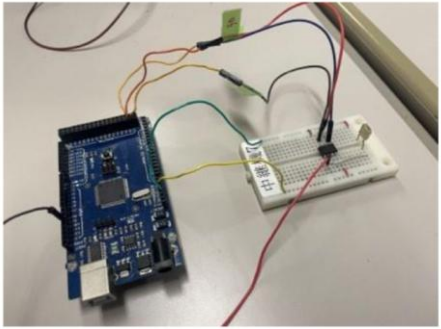 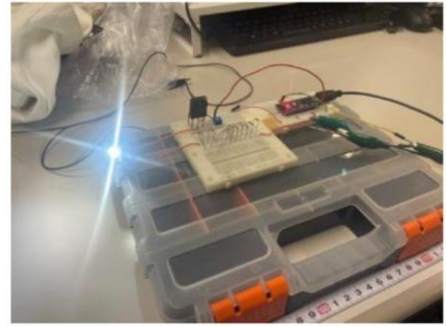

Figure 11: Experimental devices (the receiver and the transmitter)

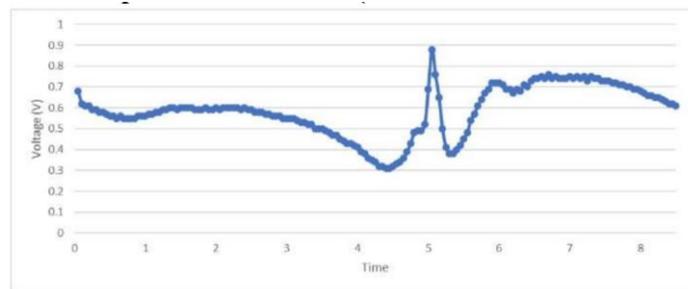

Figure 12: the sampled interference waveform stored during the idle state.

- Light source: 3 W LED
- Power transistor: C5198

**Receiver:**

- Microcontroller: Arduino Mega 1280 (ATmega1280-16AU)
- Photodiode: S6775
- AD converter: MCP3002

The actual transmitter and receiver are shown in Figure 11. To evaluate the subtraction effect, we employ On-Off Keying (OOK) modulation at the transmitter, without considering flicker, so that the performance of noise reduction can be directly observed.

Figure 12 shows the sampled interference waveform stored during the idle state. Compared with Figure 10, this confirms that one full of the periodic interference is correctly sampled by using the distortion of the waveform as a. The sampled waveform containing both signal and interference is shown in Figure 13, and the result after subtracting the interference waveform is shown in Figure 14. The resulting waveform after subtraction closely resembles a rectangular wave, demonstrating that the system operated successfully.

However, because both subtraction and correction procedures are required, the communication speed is limited. In addition, phase synchronization is achieved by initiating subtraction immediately after sampling the interference waveform. All sampling timings are controlled by a timer function, which cannot provide resolution finer than the microsecond level. Since the interference frequency in the experimental environment is 120 Hz (corresponding to a period of 8.3 ms), phase misalignment gradually accumulated across cycles. As shown in Figure 13, when such phase errors occur, excessive or insufficient subtraction can distort the waveform, preventing it from maintaining a clean rectangular shape. Therefore, correction of the subtraction values is necessary.

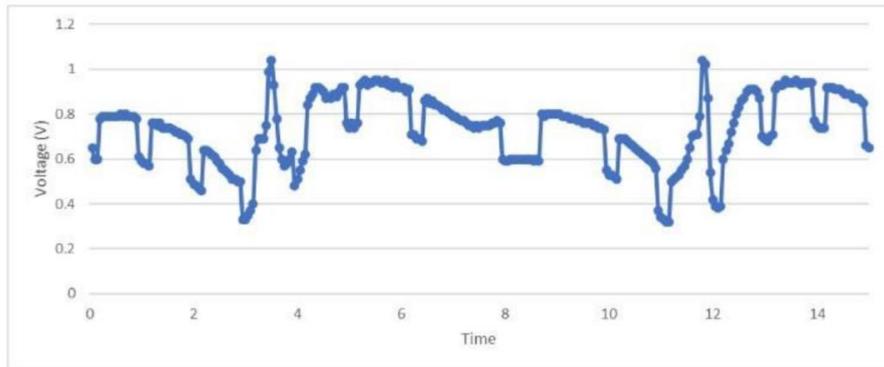

Figure 13: The sampled waveform containing both signal and interference

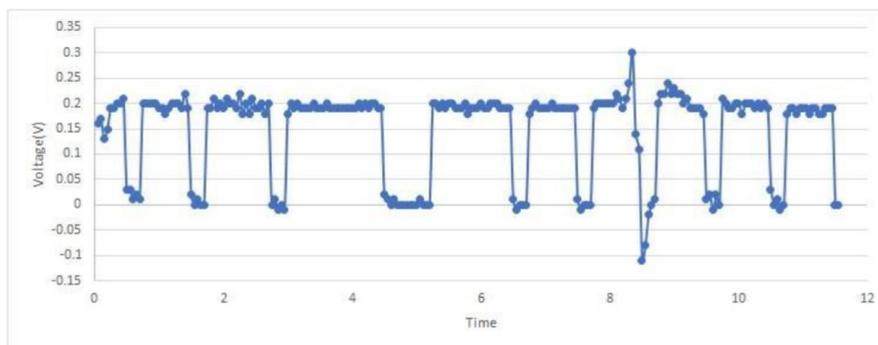

Figure 14: the result after subtracting the interference waveform

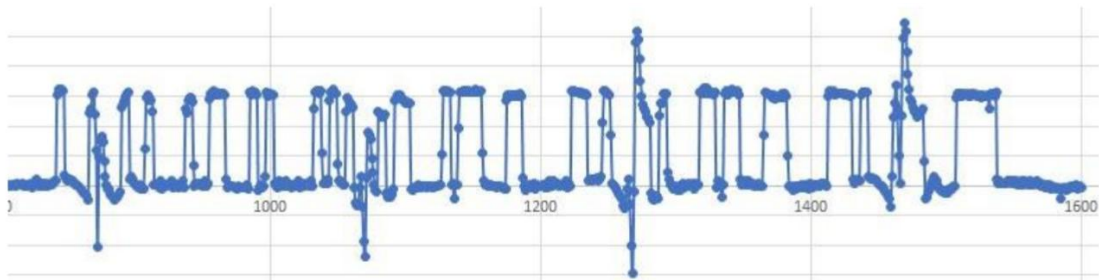

Figure 15: The waveform without reference correction, the subtracted signal in the second and subsequent cycles appeared.

### 4.2.1. Correction for HIGH signals

The mean value of the HIGH level obtained after the first cycle of subtraction is used as the reference. Because the true HIGH level is unknown to the receiver, a subtracted signal that falls below this reference may in fact correspond to a LOW symbol. Therefore, any signal above the reference is assumed to be HIGH, and correction is applied accordingly. When the subtracted signal exceeds the reference, it is considered to indicate insufficient subtraction, and the correction is applied Eq. (8):

new subtraction value
= current subtraction value + (subtracted signal − reference value). (8)

### 4.2.2. Correction for LOW signals

Since the expected level for a LOW symbol is zero, zero is used as the reference. If the subtracted signal exceeds zero, it may actually correspond to a HIGH symbol. Conversely, signals below zero is assumed to represent LOW, and correction is applied. When the subtracted signal becomes negative, it is considered an over subtraction, and the correction is performed using Eq. (9):

new subtraction value = current subtraction value + (signal value). (9)

Without correction, the subtracted signal in the second and subsequent cycles appeared as shown in Figure 15. With correction, the waveform becomes as shown in Figure 16. A comparison of these results shows that the corrected waveform more closely approximates a rectangular shape, confirming the effectiveness of the correction process.

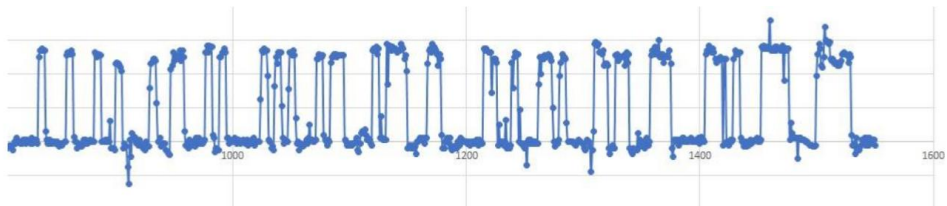

Figure 16: The waveform with reference correction

### 4.3. Experiments

In this section, we describe the performance evaluation experiments.

### 4.3.1. Experimental Conditions

The performance is evaluated using the bit error rate (BER), which is defined as the ratio of incorrectly received bits to the total number of transmitted bits. A lower BER indicates better performance. The conventional method refers to the case where no noise cancellation module is used.

The hardware setup described in Section 4.2 is used without modification. The communication distance is varied in order to change the received signal strength, and BER is measured at each distance. As an indicator of signal strength, we employ the energy-per-bit to noise density ratio ($E_b/N_0$).

International Journal of Computer Networks & Communications (IJCNC) Vol.18, No.1, January 2026

The transmitted data consist of all 256 possible 1-byte binary patterns, covering the range from 0 to 255. The receiver remains fixed in position so that the same interference waveform is present throughout all measurements, while the transmitter is moved to vary the distance. The communication speed is set to 4.0 kbps. For each measurement point, the receiver collects 20,480 bits as one measurement set. Ten sets ware collected at each point, and the average BER is calculated.

Because the transmitted data volume is large, the data are divided into segments. At the start of each segment, a trigger signal is sent from the receiver to the transmitter via a wired connection to synchronize transmission. The required data length, assuming a BER of 0.001, is calculated so that the confidence level of the experimental results is at least 0.99 (Eq. (10)):

$$\text{confidence level} = 1 - (1-p)^N, \quad (10)$$

where $p$ is the probability and $N$ is the number of samples. A photograph of the experimental setup is shown in Figure 17.

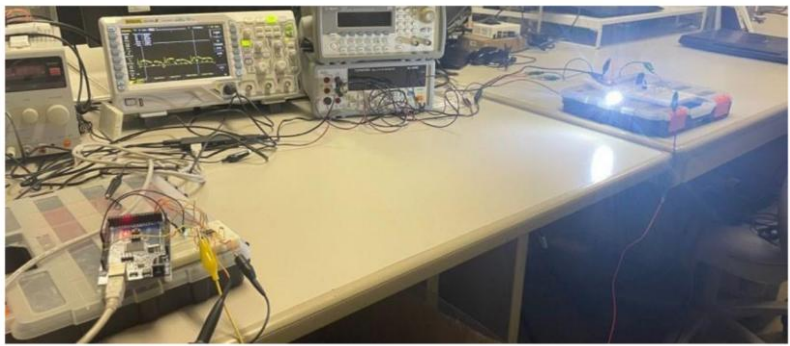

Figure 17: A photograph of the experimental setup

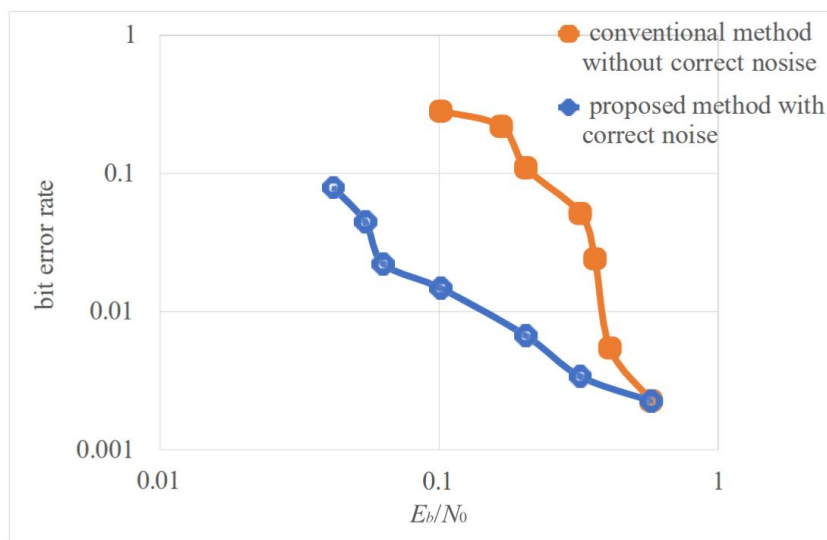

Figure 18: The experimental results with and without correct noise.

International Journal of Computer Networks & Communications (IJCNC) Vol.18, No.1, January 2026

Table 4: the result of the conventional method without noise correction.

| distance (cm) | 50 | 60 | 65 | 70 | 90 | 100 | 130 |
|---|---|---|---|---|---|---|---|
| $E_b/N_0$ | 0.575 | 0.410 | 0.362 | 0.321 | 0.205 | 0.167 | 0.102 |
| BER | 0.00227 | 0.00547 | 0.0241 | 0.0514 | 0.110 | 0.219 | 0.283 |

Table 5: the result of the proposed method with noise correction.

| distance (cm) | 50 | 70 | 90 | 130 | 210 | 230 |
|---|---|---|---|---|---|---|
| $E_b/N_0$ | 0.575 | 0.321 | 0.205 | 0.102 | 0.054 | 0.042 |
| BER | 0.00226 | 0.00339 | 0.0668 | 0.0147 | 0.0446 | 0.0791 |

### 4.3.2. Experimental Results and Discussion

The experimental results are summarized in Table 4 and Table 5, and plotted in Figure 18. In this figure, the orange curve represents the conventional method, while the blue curve represents the proposed method.

From the results, it can be observed that the influence of interference becomes significant when $E_b/N_0$ falls below 0.5. At these measurement points, the proposed method achieved better performance than the conventional method.

In the conventional method, there were points where the BER increased sharply. One possible cause is that the amplitude of the HIGH signal decreased. When this amplitude fell below the negative swing of the interference, signals that had previously been detected as HIGH were misclassified as LOW. Another factor is that the threshold was set at the approximate midpoint of the waveform. When the maximum amplitude decreased while the minimum remained unchanged, the threshold shifted, causing some signals that should have been Low to exceed the threshold and be misclassified as HIGH. Since the interference waveform is relatively smooth compared to the digital signal, such effects can cause many HIGH signals to be misclassified as LOW at certain points.

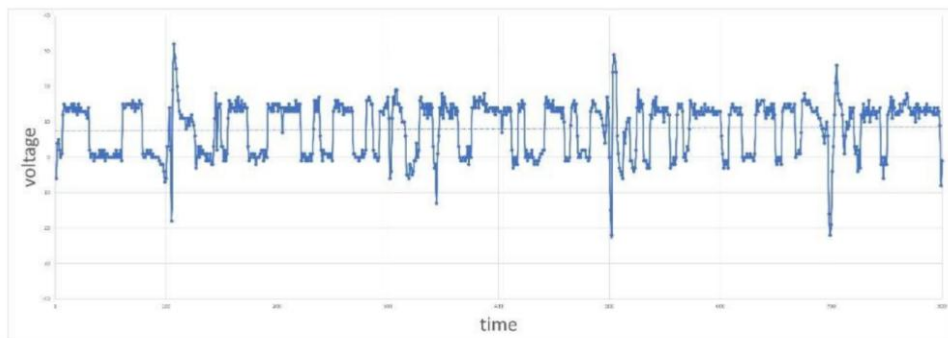

Figure 19: around Eb/N0 = 0.05, residual phase misalignment

In contrast, the BER increase in the proposed method followed a more gradual, quadratic trend. As shown in Figure 19, around $E_b/N_0$ = 0.05, residual phase misalignment in the subtraction process degraded performance. Although correction of subtraction values mitigated the impact of phase drift, it was not fully effective against high-frequency components. Nevertheless, because the BER increased more gradually than in the conventional method, the proposed approach was considered effective.

### 4.4. Conclusion

In this section, we proposed a method to mitigate the effect of interference from other light sources in VLC. By subtracting the interference waveform data from the received signal, the distorted waveform can be transformed into a shape closer to an ideal rectangular wave. Experimental evaluation of BER demonstrates that the proposed method effectively reduces the impact of interference. However, this method specifically targets periodic interference, such as that caused by fluorescent lamps powered by AC mains. As a result, it is not applicable to non-periodic optical noise. We discuss this topic at the next section.

## 5. OPTICAL NOISE MITIGATION IN VISIBLE LIGHT COMMUNICATION

In this section, we propose a solution to suppress signal distortion caused by optical noise using multiple photodetectors and subtraction circuits. The type of optical noise considered here is the same as in Section 4, namely noise generated by fluorescent lamps.

### 5.1. Active Noise Control

For noise reduction, we draw inspiration from Active Noise Control (ANC) [13], a well-known technology used in audio devices such as earphones and headphones. ANC works by superimposing an inverse-phase signal onto the noise, thereby cancelling the sound pressure. There are two main categories of control strategies, described below.

#### 5.1.1. Feedforward Control

One of the simplest implementations is the single-channel feedforward ANC system, which consists of a reference sensor to detect the incoming noise, an error sensor to monitor the residual signal, and a loudspeaker to generate the anti-noise. Feedforward systems are effective against broadband noise and are relatively easy to implement.

The reference sensor first detects the incoming noise. The noise control filter then processes this signal to generate an inverse-phase waveform, which is output through the secondary source (loudspeaker) as pseudo-noise. Finally, the noise and pseudo-noise cancel each other out, while the error sensor monitors the residual noise. The coefficients of the noise control filter are adaptively adjusted so that the error signal is minimized.

### 5.2. Proposed Method

In digital VLC systems, the photodiode used for signal reception inevitably detects both the communication signal and optical noise simultaneously. To address this issue, we focus on the feedforward control strategy described in the previous subsection, which uses a reference sensor to detect noise. Specifically, we propose a receiver design that employs an additional photodiode dedicated solely to detecting optical noise.

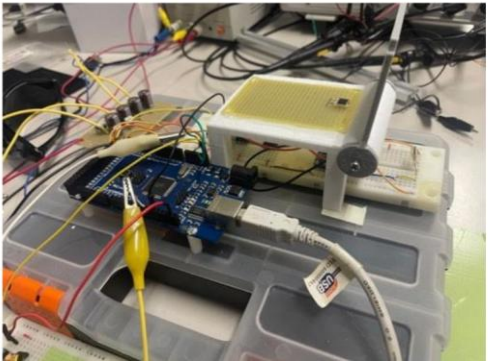
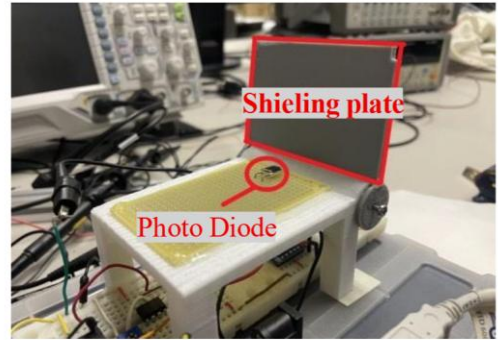

Figure 20: The actual prototype    Figure 21: The fabricated components

This design is inspired by the role of the reference sensor in feedforward ANC systems. The dedicated noise photodiode functions analogously to the reference sensor in ANC, while the subtraction circuit plays a role similar to that of the secondary source (loudspeaker), which cancels noise in audio systems. The subtraction circuit processes the waveform detected by the noise photodiode and subtracts it in real time from the waveform detected by the signal photodiode. The objective is to attenuate the effect of optical noise and recover a cleaner communication signal.

### 5.3. Implementation

We implement the proposed system and verify its operation. As in Section 4, the purpose of this experiment is to confirm the signal processing behavior. Therefore, transmission information is preloaded into the modulation microcontroller, while the demodulation microcontroller is used for demodulation and data reception. The same transmitter setup as in Section 4 is used. The actual prototype is shown in Figure 20.

The major components of the receiver are listed below:

- Microcontroller: Arduino Mega 1280
- Photodiode: S6775
- AD Converter: MCP3002
- Operational Amplifier: AD8630ARUR
- Comparator: LM358N

Since the noise photodiode is oriented toward the room lighting, it is expected to receive stronger optical noise waveforms than the signal photodiode. To control the amount of noise received, a shielding plate and a mounting stand for the noise photodiode are fabricated using a 3D printer. The fabricated components are shown in Figure 21. The shielding plate is designed with an adjustable angle, enabling regulation of the optical noise received by the noise photodiode. This mechanism is conceptually similar to the noise control filter in feedforward ANC systems.

The experimental environment included fluorescent lamp noise, as described in Section 3, producing a waveform similar to that shown in Figure 5. An example of the obtained waveforms is presented in Figure 22. In this figure, the yellow waveform corresponds to the signal photodiode output, the blue waveform corresponds to the noise photodiode output, and the purple waveform represents the result after subtraction by the noise cancellation circuit.

Table 6: The performance of our proposed method from 90cm to 170cm

| Communication distance [cm] | 90 | 110 | 130 | 150 | 170 |
|---|---|---|---|---|---|
| $E_b/N_0$ | 0.136 | 0.0785 | 0.0468 | 0.0283 | 0.0165 |
| BER | 0.000875 | 0.00575 | 0.0117 | 0.0803 | 0.297 |

It can be observed that the blue waveform still contains a small portion of the signal, indicating that part of the original signal is attenuated. This effect is likely caused by light penetration through the shielding plate, which is fabricated from PLA filament using a 3D printer. Another possible factor is reflection of the signal light from surrounding walls not covered by the shielding plate. In this paper, however, we do not further investigate this issue, since the output waveform after subtraction still approximated a rectangular signal. Additionally, small overshoots are observed in the subtracted waveform, particularly in the high-frequency components. This is attributed to sensitivity differences between the two photodiodes.

### 5.4. Experiments

In this section, we describe the performance evaluation experiments for the proposed method.

### 5.4.1. Experimental Conditions

As in Section 4, the performance is evaluated by comparing the Bit Error Rate (BER) of the proposed method with that of the conventional method without noise cancellation. The data obtained for the conventional method in Section 4 are used as the baseline for comparison.

The experimental setup consists of the receiver implemented in Section 5.2 and the transmitter described in Section 4. As before, the communication distance is varied in order to change $E_b/N_0$, and BER is measured at each point. The transmitted data consists of 256 different 1-byte patterns (0-255). The receiver is fixed in position, and the communication speed is set to 4 kbps. At each measurement point, the receiver collects 20,480 bits as one measurement set, and ten sets are collected to calculate the average BER. Because the total data size is large, transmission is divided into 80 segments. For each segment, a start signal is sent from the receiver to the transmitter via a wired connection to synchronize transmission. A photograph of the experimental setup is shown in Figure 23.

### 5.4.2. Experimental Results and Discussion

The results are summarized in Table 6 and plotted in Figure 24. In this figure, the orange curve represents the conventional method, while the blue curve represents the proposed method. As shown in Figure 24, the proposed method outperforms the conventional method at all measurement points.

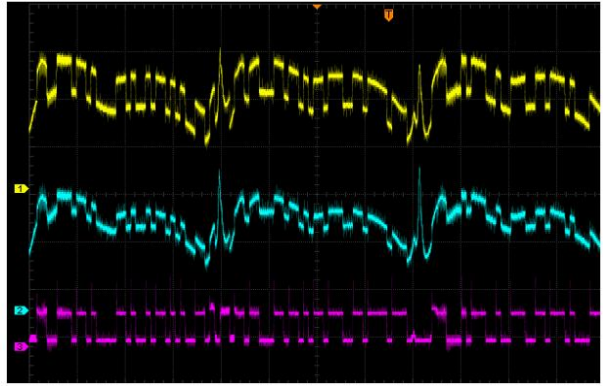

Figure 22: An example of the obtained waveforms

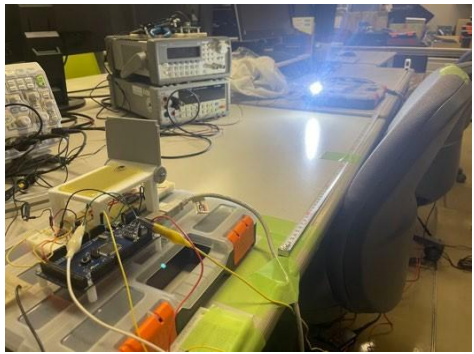

Figure 23: The photograph of the experimental setup

This improvement is attributed to the ability of the proposed device to attenuate noise waveforms, thereby reducing the impact of optical noise compared with the conventional system. The BER of the proposed method increased in a cubic trend as $E_b/N_0$ decreased. This increase is considered to result from signal attenuation at longer communication distances, which amplified the influence of high-frequency noise components that could not be fully canceled.

It is also observed that communication becomes infeasible when $E_b/N_0$ drops below 0.0165. This is likely because the HIGH-level signal voltage after subtraction becomes too small to be distinguished from LOW using the threshold. Possible countermeasures include adjusting the sensitivity of the photodiode by changing resistor values in the circuit or completely shielding the signal light from entering the noise photodiode. Compared with the method proposed in Section 4, the present method archives superior results at all measurement points. This is because, unlike the previous approach, it does not require phase adjustment of the subtraction signal using the microcontroller.

### 5.5. Conclusion

In this section, we proposed a noise reduction system for VLC inspired by Active Noise Control (ANC). By introducing a dedicated noise photodiode and performing real-time subtraction of the noise waveform from the signal waveform, the effect of optical noise can be effectively suppressed. Experimental BER measurements demonstrate that the proposed method significantly reduces the influence of interference from other light sources. However, the proposed system requires manual adjustment of the noise input, which limits its practicality. Addressing this issue remains an important direction for future work.

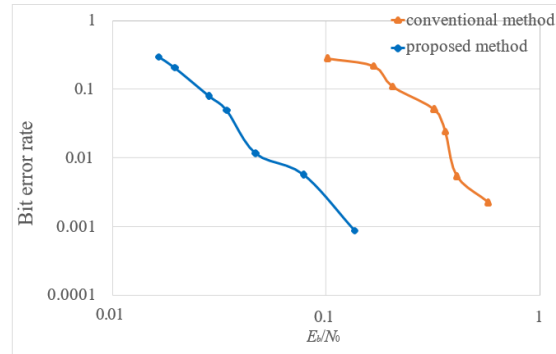

Figure 24: The communication performances from 90cm to 170 cm.

## 6. CONCLUSIONS

In this paper, we focused on the problem of signal distortion in visible light communication (VLC) caused by optical noise from ambient light sources and proposed two mitigation methods based on noise data.

In Section 4, we proposed a method in which interference waveforms were pre-acquired and used by a microcontroller to modify the received signal. BER measurements were conducted for both a conventional receiver without noise reduction and a receiver implementing the proposed method. The results demonstrated that the proposed approach was effective, particularly under low $E_b/N_0$ conditions. In Section 5, we proposed another method using a dedicated noise photodiode and a subtraction circuit to attenuate optical noise in real time. Again, BER measurements confirmed that this approach effectively reduced the influence of noise compared to the conventional method.
Each method has its advantages and limitations. The interference cancellation method presented in Section 4 features a simple receiver circuit that is easy to implement but requires more complex microcontroller processing and is less effective against high-frequency noise components. On the other hand, the noise-cancelling method presented in Section 5 simplifies microcontroller processing but requires manual adjustment of the electrical signal obtained from the noise photodiode, which limits its practicality. Therefore, an important subject for future work will be to address these limitations—developing a method that combines simple implementation, effective suppression of high-frequency noise, and automated adjustment of the noise input.

## ACKNOWLEDGEMENTS

This research was partially supported by the Research Program of Ryukoku University.

## REFERENCES

[1]     S. Haruyama, "Visible light communication," Journal of IEICE, 94(12) - D, pp. 1055–1059, 2011.
[2]     R. Sagotra and R. Aggarwal, "Visible Light Communication," International Journal of Computer Trends and Technology (IJCTT) Vol. 4, No. 4, pp. 906–910, 2013.

[3] D. C. O'Brien, at. el. , "Visible light communications: Challenges and possibilities," 2008 IEEE 19th International Symposium on Personal, Indoor and Mobile Radio Communications, Cannes, 2008, pp. 1–5.
[4] T. Saito, "A Study for flicker on Visible Light Communication," Technical Report of IEICE CS, Vol. 106, No. 450, pp. 31–35, 2007.
[5] I. Shouichi, "Reduction of Flicker by Coding and Modulation for Visible-Light Communication," Technical Report of IEICE OCS, Vol. 108, No. 39, pp. 1–4, 2008.
[6] K. Okuda, R. Yoneda, T. Nakamura, and W. Uemura, "A Warning System for Overspeed at the Corner Using Visible Light Based Road-To-Vehicle Communication," International Journal of Ad hoc, Sensor and Ubiquitous Computing (IJASUC), Vol. 6, pp. 1–9, (2015).
[7] W. Uemura and Takahiro Kitazawa, "A Hybrid Modulation Method for Dimming in Visible Light Communication," International Journal of Computer Networks and Communications (IJCNC), Vol. 10, pp. 51–59, (2018).
[8] W. Uemura and K. Shimizu, "A Selecting Robots Method Using Visible Light Communication," International Journal of Ad hoc, Sensor and Ubiquitous Computing (IJASUC), Vol. 9, pp. 1–8, (2018).
[9] W. Uemura, Y. Fukumori, and T. Hayama, "About Digital Communication Methods for Visible Light Communication," International Journal of Computer Networks and Communications (IJCNC), Vol. 13, pp. 1–13, (2021).
[10] Y.Ashida, K.Okuda and W.Uemura. "A VLC receiving devise using audio jacks with a folding noise," 2012 International Symposium on Information Theory and its Applications, 475–479, (2012).
[11] K. Okuda, T. Yamamoto, T. Nakamura, and W. Uemura, "The Key Providing System for Wireless Lan Using Visible LightCommunication," International Journal of Ad hoc, Sensor and Ubiquitous Computing (IJASUC),Vol. 5, pp. 13–20, (2014).
[12] K. Okuda, H. Shirai, T. Nakamura, and W. Uemura, "A Novel Keyless Entry System Using Visible Light Communication," International Journal of Ad hoc, Sensor and Ubiquitous Computing (IJASUC),Vol. 5, pp. 1–8, (2014).
[13] T. Hamano and W. Uemura, "Effect Reduction Method by Periodic Noise on Pulse Wave Visible Light Communication," 2022 IEEE 11th Global Conference on Consumer Electronics (GCCE), Osaka, Japan, 2022, pp. 661-663, doi: 10.1109/GCCE56475.2022.10014310.

## AUTHORS

**Wataru Uemura** was born in 1977, and received B.E, M.E. and D.E. degrees from Osaka City University, in 2000, 2002, and 2005. He is an associate professor of the Faculty of Advanced Science and Technology, Ryukoku University in Shiga, Japan. He is a member of IEEE, RoboCup and others. He is a chairperson of RoboCup Japanese Regional Committee.

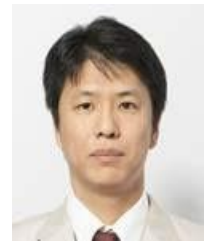

**Takumi Hamanoi** was born in 1998 and received B.E, and M.E from Ryukoku University, in 2021, 2023. He is interested in visible light communication

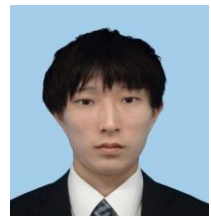